\documentclass[12pt,preprint]{aastex}

\shortauthors{Okamoto et al.}
\shorttitle{EMERGING HELICAL FLUX ROPE UNDER PROMINENCE}

\begin{document}

\title{EMERGENCE OF A HELICAL FLUX ROPE UNDER AN ACTIVE REGION PROMINENCE}
\author{\textsc{
Takenori J. Okamoto,$^{1,2}$\footnote{T. J. O. is supported by the Research Fellowships from the Japan Society for the Promotion of Science for Young Scientists.}
Saku Tsuneta,$^1$
Bruce W. Lites,$^3$
Masahito Kubo,$^3$
Takaaki Yokoyama,$^4$
Thomas E. Berger,$^5$
Kiyoshi Ichimoto,$^1$
Yukio Katsukawa,$^1$
Shin'ichi Nagata,$^2$
Kazunari Shibata,$^2$
Toshifumi Shimizu,$^6$
Richard A. Shine,$^5$
Yoshinori Suematsu,$^1$
Theodore D. Tarbell,$^5$
and Alan M. Title$^5$
}}
\affil{
$^{1}$National Astronomical Observatory, Mitaka, Tokyo, 181-8588, Japan\\
$^{2}$Kwasan and Hida Observatories, Kyoto University, Yamashina, Kyoto, 607-8471, Japan\\
$^{3}$High Altitude Observatory, National Center for Atmospheric Research, P.O. Box 3000, Boulder CO 80307-3000, USA%
\footnote{The National Center for Atmospheric Research is sponsored by the National Science Foundation.}\\
$^{4}$Department of Earth and Planetary Science, School of Science, University of Tokyo, Hongo, Bunkyo, Tokyo, 113-0033, Japan\\
$^{5}$Lockheed Martin Solar and Astrophysics Laboratory, B/252, 3251 Hanover St., Palo Alto, CA 94304, USA\\
$^{6}$ISAS/JAXA, Sagamihara, Kanagawa, 229-8510, Japan\\ 
\ \\
{\it Received} 2007 October 27; {\it accepted} 2007 December 18
}
\email{joten.okamoto@nao.ac.jp}

\begin{abstract}

Continuous observations were obtained of active region 10953 with the Solar Optical Telescope (SOT) on board the \emph{Hinode} satellite during 2007 April 28 to May 9.
A prominence was located over the polarity inversion line (PIL) in the south-east of the main sunspot.
These observations provided us with a time series of vector magnetic fields on the photosphere under the prominence.
We found four features:
(1) The abutting opposite-polarity regions on the two sides along the PIL first grew laterally in size and then narrowed.
(2) These abutting regions contained vertically-weak, but horizontally-strong magnetic fields.
(3) The orientations of the horizontal magnetic fields along the PIL on the photosphere gradually changed with time from a normal-polarity configuration to a inverse-polarity one.
(4) The horizontal-magnetic field region was blueshifted.
These indicate that helical flux rope was emerging from below the photosphere into the corona along the PIL under the pre-existing prominence.
We suggest that this supply of a helical magnetic flux into the corona is associated with evolution and maintenance of active-region prominences.

\end{abstract}


\section{Introduction}

Solar prominences are cool material (10$^4$ K) floating in the corona (10$^6$ K) and located over polarity inversion lines (PILs) of the photosphere.
It is known that prominences are supported by coronal magnetic fields against gravity (see references in Martin 1998).
Such magnetic fields have dipped shapes, and the cool plasma is sitting at the bottom of the dips 
(Kippenhahn \& Schl\"{u}ter 1957).
Moreover, many observations of prominences show inverse polarity (Leroy et al. 1984; Tandberg-Hanssen 1995; Lites 2005),
which means that the direction of horizontal magnetic fields is toward the positive polarity side from the negative polarity side.
Hence, it is often thought that magnetic fields in prominences have helical structures 
(e.g., Kuperus \& Tandberg-Hanssen 1967; Kuperus \& Raadu 1974; Hirayama 1985).
This topology of magnetic fields is suggested from observations of erupting prominences, or coronal mass ejections
(e.g., Dere et al. 1999; Ciaravella et al. 2000; Low 2001).

How is such a helical magnetic field created in association with prominences in the corona ?
Two theories have been discussed: flux rope models (e.g., Rust \& Kumar 1994; Low \& Hundhausen 1995; Low 1996; Lites 2005; Zhang \& Low 2005) and sheared-arcade models 
(e.g., Pneuman 1983; van Ballegooijen \& Martens 1989, 1990; Antiochos et al. 1994; DeVore \& Antiochos 2000; Martens \& Zwaan 2001; Aulanier et al. 2002; Karpen et al. 2003;
Mackay \& van Ballegooijen 2005, 2006).
In the former model, an originally-twisted flux rope emerges from below the photosphere into the corona.
The helical structure is thought to be made by the convection in the solar interior.
In the latter one, potential fields in the corona are sheared by the photospheric motion along PILs.
As a result, magnetic reconnection occurs among the sheared fields, and then helical fields are constructed in the corona.
Both models have the same final configuration, but the processes leading to it are quite different.
However, we have no clear observational evidence to support either model, although emerging twisted magnetic fields related to flares have already been observed
(e.g., Tanaka 1991; Lites et al. 1995; Leka et al. 1996; Ishii et al. 1998).
We have had difficulties due to seeing and discontinuity of observing time with ground-based observations in revealing the mechanism of prominence formation.

The \emph{Hinode} satellite (Kosugi et al. 2007) has a sun-synchronous polar orbit, so that we can observe the Sun continuously without interruption by a spacecraft night.
The Solar Optical Telescope (SOT, Tsuneta et al. 2007; Suematsu et al. 2007; Ichimoto et al. 2007; Shimizu et al. 2007) on board \emph{Hinode}
has the Spectro-Polarimeter (SP), which provides long sequences of vector magnetic field measurements at high spatial resolution.
We have an observation of one active region that had a pre-existing prominence to investigate the evolution of its associated magnetic fields on the photosphere.
We report on the results and findings from this observation in this letter.

\section{Observation and Data Analysis}

The \emph{Hinode} satellite was tracking a sunspot in active region NOAA 10953 from 2007 April 28 to May 9.
At the beginning of this period, we had periodic scannings of this active region including a pre-existing prominence 
with the SP as well as time series of images with the Filtergraph (FG).
The multiple scanning was mainly performed with Fast Mapping mode,
which has an integration time of 3.2 seconds for one slit and the pixel sampling of 0''.32.
The average cadence of the scanning was three hours and the field of view was 160''$\times$160" (512$\times$512 pixels).
The SP simultaneously measured the full Stokes profiles of the Fe I lines at 6301.5 and 6302.5 \AA\ with a sampling of 21.6 m\AA.

Vector magnetic fields were derived from the calibrated Stokes profiles on the assumption of a Milne-Eddington atmosphere.
We resolved the 180$^{\circ}$ azimuth ambiguity with the AZAM utility (Lites et al. 1995),
where the basic premise is minimization of spatial discontinuities in the field orientation.

Dopplergrams were also produced by the inversion.
However, we cannot determine the absolute zero velocity since these include various kinds of Dopplershift components.
Therefore, we define the average velocity in the plage region for which vertical field strength is larger than 1000 Gauss as zero.

  \begin{figure}
\epsscale{0.7}
 \plotone{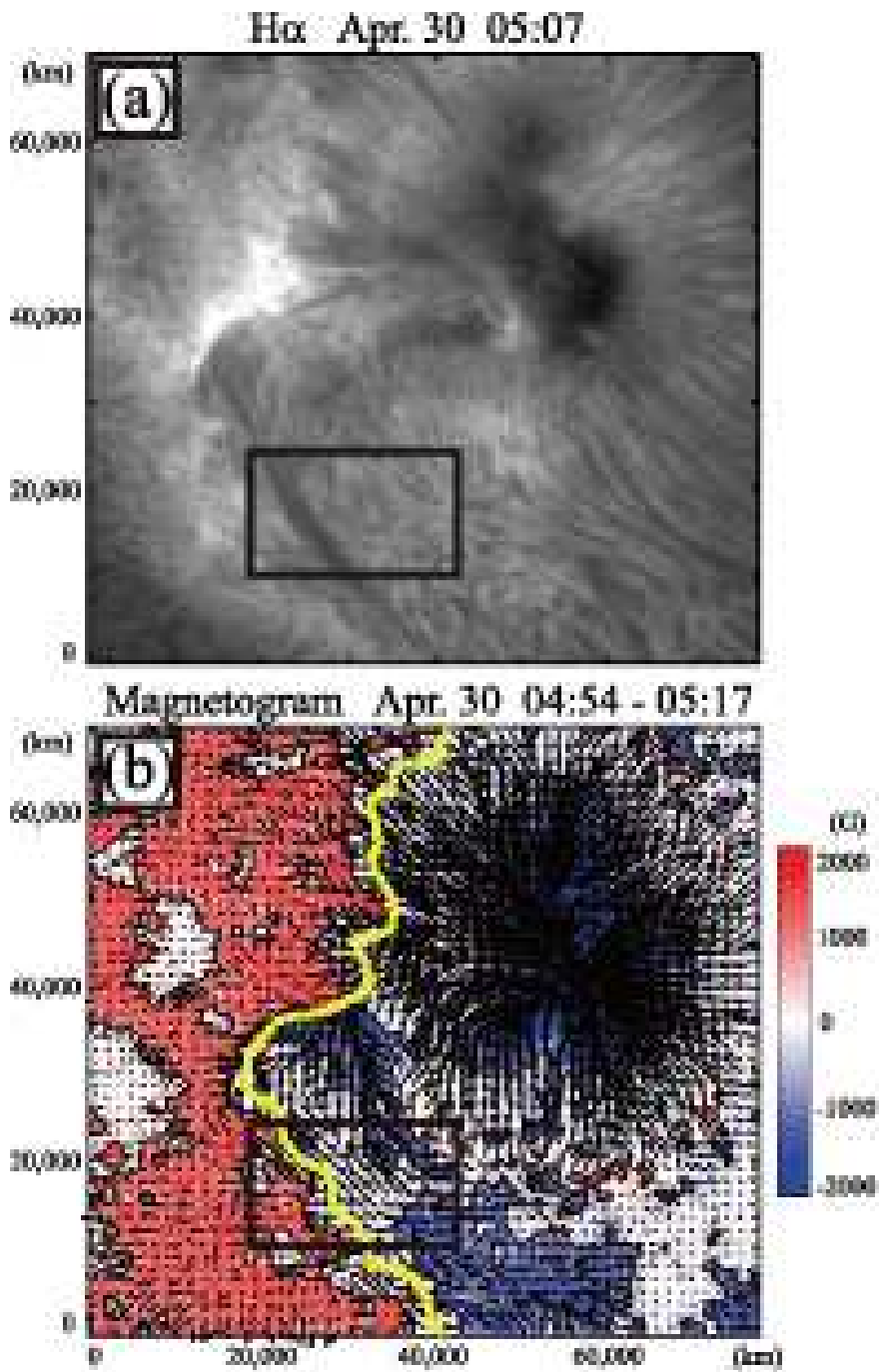}
    \caption{
({\it a}) H$\alpha$ image taken with the FG of \emph{Hinode}/SOT. North is up and east is to the left.
({\it b}) Magnetogram derived from the SP. The field of view is the same as that of Panel {\it a}. 
The scanning was performed from east to west during 04:51 to 05:23 UT in this field of view.
Color contour indicates the strength of vertical magnetic fields; red is positive and blue is negative.
Arrows show the strength of horizontal magnetic fields.
The yellow line is PIL, which is calculated from the smoothed vertical magnetic fields.
The black line contour means $\pm$650 Gauss boundary of the vertical magnetic fields.
The vector magnetic fields are converted in the solar local frame.
}
    \label{fig1}
  \end{figure}

\section{Result}

Figure \ref{fig1} shows an H$\alpha$ filtergram, and a vector magnetogram of the active region taken or scanned around 05:00 UT on 2007 April 30.
The magnetogram displays vertical magnetic field strength (color contour; red is positive and blue is negative),
horizontal magnetic field strength (arrow), the PIL (yellow line), and the $\pm$650 Gauss boundary of vertical fields (black line contour).
The vector magnetic fields are transformed into the solar local frame.
A prominence in the south-east of the sunspot was located along the PIL.
We notice that horizontal magnetic fields under the prominence have the same direction.
The direction is almost parallel to the prominence.
Moreover, the opposite-polarity regions on the both sides of the PIL have mainly horizontal fields, as indicated by the relatively weak vertical magnetic field (Fig. \ref{fig1}b).
Hereafter, we call this region the {\it weak-field region}.
The reason we drew the 650-G boundary for the vertical fields is that this value is the averaged strength of the horizontal magnetic field near the PIL.
The 650-G boundary appears to be the boundary separating the outer vertical-field dominated region from the inner horizontal-field dominated region although this is less clear for sunspot side.

\begin{figure}
\epsscale{1.0}
\rotatebox{90}{
\plotone{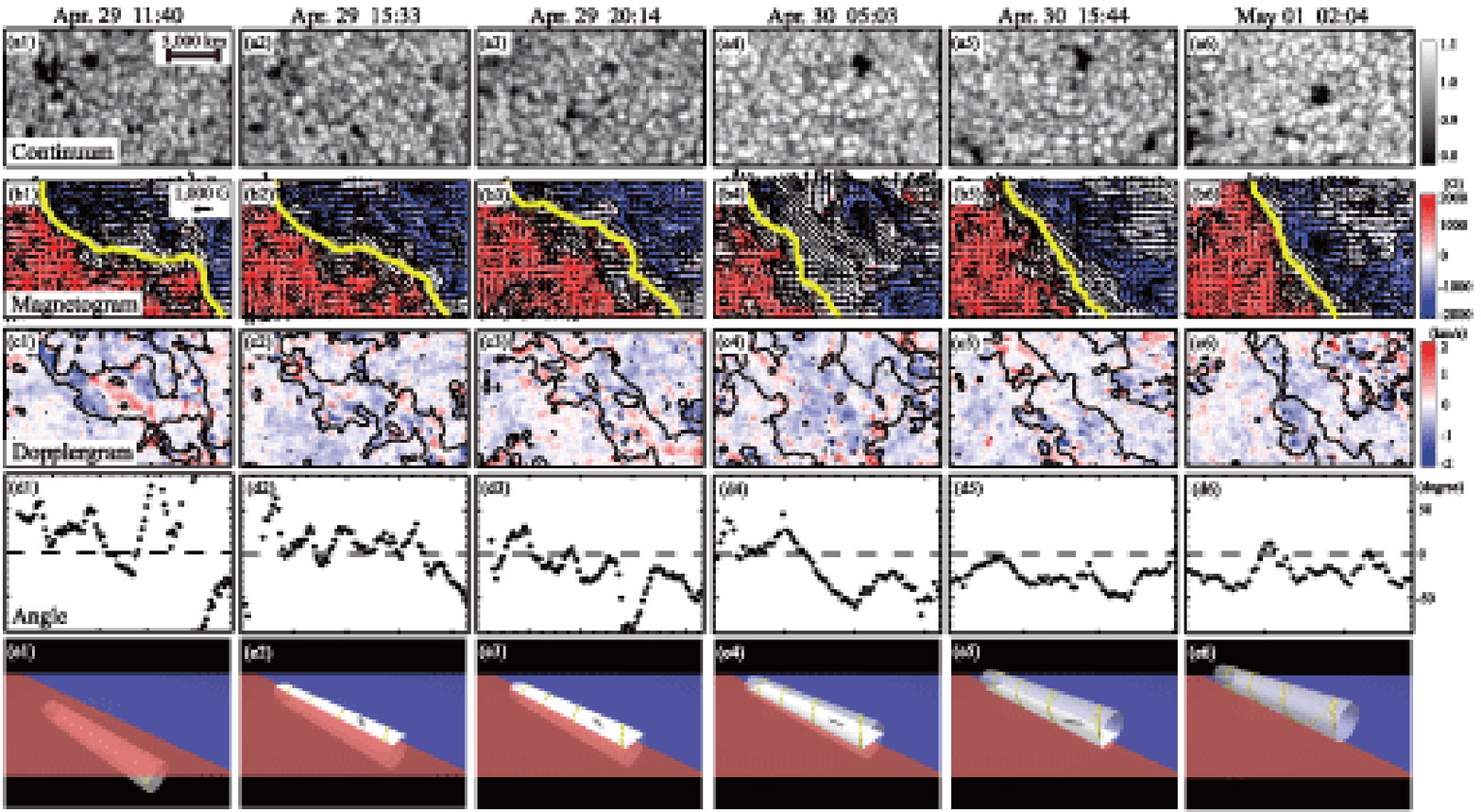}
}
\caption{Time series of SP data. Panel ({\it b4}) is the same as the area indicated by the black solid box in Figure \ref{fig1}. 
The field of view is 32''$\times$19'' (23,000 km$\times$14,000 km).
The scanning time is about six minutes in these small regions and the middle time is displayed at the top of each column.
({\it a1--a6}) Continuum images.
({\it b1--b6}) Magnetograms: Same as Figure \ref{fig1}{\it b}.
({\it c1--c6}) Dopplergrams: Red (blue) indicates redshift (blueshift). Black lines are the same as those of {\it b1--b6}.
({\it d1--d6}) Angles between the PIL and the orientations of horizontal magnetic fields: Positive (negative) means ``normal (inverse) polarity''.
({\it e1--e6}) Schematic interpretation of our observations: The red (blue) region indicates the plage (sunspot side) on the photosphere. The white tube is a helical flux rope.
The yellow line shows a magnetic field line on the surface of the flux tube.
Black arrows indicate the orientation of horizontal magnetic fields of this rope crossing with the photosphere.
}
    \label{fig2}
  \end{figure}

We focus on the evolution of the photospheric magnetic field along the PIL.
Panels {\it a1--a6}, {\it b1--b6}, and {\it c1--c6} in Figure \ref{fig2} show time series of continuum images, vector magnetograms, and Dopplergrams, respectively.
The field of view is indicated with the black solid box in Figure \ref{fig1}.
In the series of the magnetograms, we found that the {\it weak-field region} broadened (Panels {\it b2--b4} in Fig. \ref{fig2}) and then narrowed (panels {\it b5--b6} in Fig. \ref{fig2}).
In other words, the abutting opposite polarities were spatially widened, and closed like a sliding door.
The {\it weak-field region} is blueshifted (Panels {\it c2--c6} in Fig. \ref{fig2}).
We investigated the angle between the PILs and the horizontal magnetic fields (Panels {\it d1--d6} in Fig. \ref{fig2}).
Positive angle indicates the normal polarity, while negative angle means the inverse polarity.
We find that the angles along the PILs gradually decreased from positive to negative during the widening and narrowing process of the {\it weak-field region}.

These observational results indicate emergence of a helical flux rope from below the photosphere into the corona.
The scenario is as follows.
First, a helical flux rope is located along the PIL under the photosphere (Panel {\it e1} in Fig. \ref{fig2}).
This horizontally oriented rope with weak twist explains the dominance of the horizontal fields on the two sides of the PIL as the rope rises into the photosphere.
Second, as the rope rises up, it pushes aside the occupying photospheric plasma containing vertical fields.
When the top of the rope reaches the photosphere, we observe its horizontal field on the photosphere in what is called the normal-polarity configuration (Panel {\it e2} in Fig. \ref{fig2}).
As the rope continues to rise through the photosphere, the angle between the PIL and the observed horizontal fields of the rope becomes smaller (Panel {\it e3} in Fig. \ref{fig2}).
In a cylindrically shaped twisted flux rope, we expect the field to have an azimuthal component in its outer part of the rope.
Hence, as the rope continues to rise through the photosphere, 
the angle between the PIL and the observed horizontal fields of the rope becomes smaller until the axis of the rope has reached the photosphere (Panel {\it e3} in Fig. \ref{fig2}).
Then, after the rope axis has risen above the photosphere, the angle shows the inverse polarity because the lower part of the helical rope crosses the photosphere (Panel {\it e5} in Fig. \ref{fig2}).
Simultaneously, the {\it weak-field region} narrows as the cross section decreases.
Thus, throughout the rising of the flux, a blueshift is seen in the {\it weak-field region}.

\section{Discussion}

The \emph{Hinode}/SP observations presented here offer clear evidence of an emerging helical flux rope along the PIL under an active-region prominence:
The {\it weak-field region} became first wider and then narrower during the emergence of the flux rope,
while the orientation of the horizontal magnetic fields observed at the photosphere changes from the normal-polarity to the inverse-polarity configuration.
We note that we distinguish the {\it weak-field region} from the filament channel since the {\it weak-field region} appears temporarily in the filament channel.
Cheung et al. (2007) observed similar polarity reversal in their radiative MHD simulation.
Lites (2005) pointed out in a similar evolutionary event that the abutting opposite polarities may spatially separate under an active-region prominence
although the cadence of that observation was one day and the resolution was much lower than that of ours.
Gaizauskas et al. (1997) also reported that elongated voids appeared along a filament channel seen in off-band H$\alpha$ a few days before prominence formation.
Our results with \emph{Hinode}/SOT suggest that such broadening is associated with emergence of the horizontal flux tubes.

We obtain the following physical parameters from the ME inversion of the SP data for our emerging flux rope event.
The {\it weak-field region} had a width of up to 10,000 km, a duration of one and a half days, and an upward velocity of up to 300 m $s^{-1}$.
Hence, the diameter of the flux tube is roughly estimated to be 39,000 km if the maximum upward velocity continues for one and a half days.
This estimation is consistent with both the maximum width of the {\it weak-field region} and typical height of active-region prominences on the limb.
The filling factor and strength of horizontal magnetic fields are estimated to be 0.15 and 650 Gauss on average in the {\it weak-field region}, respectively.
When this flux is supplied into the corona, the average strength of the magnetic fields is 100 Gauss.
This value is consistent with active-region prominences according to previous measurements of magnetic fields
(e.g., Tandberg-Hanssen and Malville 1974; Wiehr and Stellmacher 1991; Casini et al. 2003; Okamoto et al. 2007),
and this flux rope is strong enough to support prominences.
Therefore, we suggest that this emergence of the helical magnetic flux rope is associated with evolution and maintenance of the prominence.

This is the only observation made so far of an active-region prominence, and we do not rule out the sheared arcade model (see Introduction).
It will be useful to observe quantitatively both active-region and quiescent prominences with the SP to determine whether the event reported in this paper is common in the solar atmosphere.
We suggest that the above observational evidence must be closely related to the evolution of a prominence via the emergence of a twisted magnetic flux rope.
The H$\alpha$ prominence was already seen before the start of the emerging episode reported here.
The appearance of the pre-existing prominence was considerably changed, but after the emergence of the flux rope, the prominence became stable.
We will report the relationship between the emerging helical flux rope and the prominence activity in a subsequence paper.

\

The authors thank B. C. Low for useful comments.
\emph{Hinode} is a Japanese mission developed and launched by ISAS/JAXA, with NAOJ as domestic partner and NASA and STFC (UK) as international partners.
It is operated by these agencies in co-operation with ESA and NSC (Norway).
This work was carried out at the NAOJ Hinode science center, which was supported by the Grant-in-Aid for Creative Scientific Research ``The Basic Study of
Space Weather Prediction'' from MEXT, Japan (Head Investigator: K. Shibata), generous donation from the Sun Microsystems Inc., and NAOJ internal funding.

\end{document}